\documentclass[useAMS]{mn2e}
\usepackage{graphicx}
\usepackage{amssymb}
\usepackage{epstopdf}
\font\tenbg=cmmib10 at 10pt

\def \rvecphi{{\hbox{\tenbg\char'036}}}

\def\lesssim{\mathrel{\hbox{\rlap{\hbox{\lower4pt\hbox{$\sim$}}}\hbox{$<$}}}}
\def\gtrsim{\mathrel{\hbox{\rlap{\hbox{\lower4pt\hbox{$\sim$}}}\hbox{$>$}}}}

\title{Planet Migration and Disk Destruction due
to Magneto-Centrifugal  Stellar Winds}

\author[R.V.E. Lovelace et al.]{
R.V.E. Lovelace,$^{1,2}$
M.M. Romanova,$^{1}$ and
A.W. Barnard$^{2}$\\
$^1$ Department of Astronomy, Cornell University, Ithaca, NY
14853-6801, \\
$^2$ Department of Applied and Engineering Physics, Cornell
University, Ithaca, NY 14853-6801}

\begin{document}
\maketitle
\label{firstpage}

\begin{abstract}

    This paper investigates the influence of
magneto-centrifugally driven or simply
magnetic winds of rapidly-rotating,
strongly-magnetized T Tauri stars in causing the
inward or outward migration of close-in giant
planets. 
  The azimuthal ram pressure of the magnetized wind
acting on the planet tends
to increase the planet's angular momentum and cause
outward migration if
the star's rotation period $P_*$ is less than the planet's
orbital period $P_p$.  In the opposite case,
$P_* > P_p$, the planet migrates inward.  
   Thus, planets orbiting at distances larger (smaller)
than $0.06 {\rm AU}(P_*/5{\rm d})^{2/3}$ tend to be
pushed outward (inward), where $P_*$ is the rotation
period of the star assumed to have the mass of the sun.
    The magnetic winds are likely to occur in T Tauri
stars where the thermal
speed of the gas close to the star is small,
where the star's magnetic field
is strong, and where the star rotates rapidly. 
    The time-scale for appreciable radial motion of the
planet is estimated as $\sim 2 - 20$ Myr.  
  A sufficiently massive close-in  planet may cause tidal locking
and once this happens
the radial migration due to the magnetic wind ceases.
   The magnetic winds
are expected to be important
for planet migration for the case of a 
multipolar magnetic field rather than 
a dipole field where the
wind is directed away from the
equatorial plane and where a magnetospheric
cavity forms.  
    The influence of the magnetic wind in eroding and
eventually destroying the accretion disk is
analyzed.  A momentum integral is derived
for the turbulent wind/disk boundary layer and this
is used to estimate the disk erosion time-scale  
as $\sim 1-10^2$ Myr,
with the lower value favored.

\end{abstract}


\section{Introduction}

 More than 250 planets have been discovered around solar-type
stars. About $20\%$ of them are located close to the
star, at $R < 0.1$ AU, and a significant number are located as close
as at $R<0.05$ AU.
   According to the presently favored
interpretation, planets form far away from the star either through
core accretion (Mizuno 1980; Pollack  et al. 1996), or through
instabilities in the disk (Boss 2001).
   Subsequently, they migrate inward
due to their gravitational interaction with the disk
(reviewed by Papaloizou \& Terquem 2006).
 For typical conditions  planets migrate inward as
a result of interaction of the planet with the disk matter.
 The planet loses part of its orbital angular momentum by overtaking
collisions with the disk outside its orbit and it gains a smaller
part by overtaking collisions of the disk matter inside its orbit.

   Some planets are expected to migrate
close to the star where the disk properties
are strongly influenced by the star and its rotating
magnetic field.
   In particular, the inner regions of the disk
may be dispersed as a result of heating by the star (Kuchner
\& Lecar 2002) and photo-evaporation 
(Matsuyama, Johnstone \& Murray 2003).
  Alternatively, the equatorial-plane density of the accretion
disk may be enormously reduced inside of what
is termed the ``magnetospheric gap'' ($R<R_m$) owing to
to the strong  magnetic field of the rotating
protostar (Lin, Bodenheimer \& Richardson 1996; 
Romanova \& Lovelace 2006).
  In this situation inward migration of the planet 
stops once the planet is some distance inside
the gap ($R<0.63R_m$) because there is no matter
for it to transfer its angular momentum to. 
   On the other hand, the
magnetic interaction between a planet 
and the star may also influence
a planet's migration  
 (Papaloizou 2007; Fleck 2008).
    Indications of such interactions have been observed by
Shkolnik et al. (2005). 
    In these models,  the star's magnetic
field is assumed to be dipolar with sufficient
strength  that a low density cavity forms
near the star.

  In contrast, if the star's magnetic field 
is dominantly multipolar
rather than dipolar (e.g., Safier 1998;
Donati et al. 2007; Johns-Krull 2007), 
then there may be no cavity near the star.
   Instead the magnetic field consists of closed field
regions and multiple open field line regions where the field
extends to large distances.
   A wind can flow outward along
the open field lines (Jardine et al. 2006;
Gregory et al. 2006). 
  Spectral measurements  provide evidence of strong stellar
winds from young stars (e.g., Kwan, Edwards, \& Fischer 2007).

    The winds from
magnetized stars may be thermally
driven as the Solar wind is as proposed 
by Matt and Pudritz (2008a,b). 
   In this case, the magnetic field still has an important
role in the outward transport of angular momentum 
(Weber \& Davis 1967; Matt \& Pudritz 2008a,b).
    In contrast, for conditions where the thermal
speed of the gas close to the star is small (compared
with the escape velocity), where the star's magnetic field
is strong, and where the star rotates rapidly,
there are magneto-centrifugally driven
winds (or more simply magnetic winds)
(Michel 1969, M69 hereafter; 
Belcher \& MacGregor 1976.  BM76 hereafter). 
   The rapid rotation of the stars results of course
from the accretion of high specific angular momentum
from a disk.
    The magneto-centrifugal winds are driven by the 
star's rapidly rotating magnetic
field rather than the thermal
energy in the star's corona.
  The angular momentum and energy carried by
the wind is extracted from the star's rotation.
   The magnetic fields of classical T
Tauri stars (CTTSs) are typically 
a thousand times larger the field of the Sun (e.g. Johns-Krull
\& Valenti 2000; Donati, et al. 2007), and these stars rotate 
much faster than the Sun with
periods $\sim 2-10$ d (Bouvier et al. 1993). 
   Thus magnetically driven winds may important for T Tauri
stars.

 Magnetic stellar winds
have been found in magnetohydrodynamic (MHD)
simulations  of rapidly-rotating,
disk-accreting stars in the
``propeller" regime (Romanova et al. 2005; Ustyugova et al. 2006).
  The winds are found to flow in opposite directions
along the rotation axis transporting energy and angular
momentum (via the electromagnetic field) away from the star.
   So far simulations have been done only for the case of 
a dipole field aligned with the rotation axis.  
   In this case the  dipole magnetic field
acts to block an equatorial outflow.
   As mentioned for the case of
multipolar fields, winds are expected in all
directions from the star's surface.

   This paper first  analyzes 
the torque due to magnetic
stellar winds  on close-in giant planets.  
  This torque can cause the planet to migrate either
inward or outward depending on the star's rotation
period and the planet's orbital period.
   We go on to develop a model for 
the erosion of disks by magnetic winds. 
   It is thought that 
close-in giant planets formed at larger distances ($\gtrsim 5$ AU)
and migrated rapidly inward due to angular momentum
loss to the accretion disk (type II migration;  Papaloizou
\& Terquem 2006).     
    In addition to the influence of the
stellar wind, a number of other effects may also be important
in preventing the planet from migrating all the way into the star:  
   (1) As mentioned above,
there may  be a  magnetospheric gap near the star
where the disk density is greatly reduced so that the 
inward migration stops. 
  (2) The inner region of the disk may be dispered
by heating by the star as mentioned, and it may be
eroded by a magnetic wind.
  (3) There may be a strong tidal interaction which acts to
lock the star's surface layer to the planets motion 
(Zahn 1994; Marcy et al. 1997; Donati et al. 2007). 
  Once locking occurs the migration due
to a magnetic wind stops.

   Section 2 of this paper discusses magneto-centrifugally
stellar winds (or simply magnetic winds).   
  Section 3 analyzes the influence of
magnetic winds  in pushing  close-in planets outward or inward. 
  Section 4 comments on the influence of the tidal
interaction on close-in planet migration.
  Section 5 investigates the role 
of magnetic winds in eroding and eventually destroying the accretion
disk.
    Section 6 gives the conclusions of this work.

\section{Magneto-Centrifugally Driven Winds}

   We adopt the Weber and Davis (1976) model of
an ideal magnetohydrodynamic (MHD) 
stellar wind, where the wind is
assumed stationary and axisymmetric, and where
attention is focused on the equatorial region
of the flow.  
   In spherical $(R,~\theta,~\phi)$ inertial coordinates,
the density, pressure, and flow velocity of the wind are
$\rho(R)$, $p(R)$, and
${\bf v}(R)=v_R \hat{\bf R} + v_\phi \hat{\rvecphi~}$.
   The magnetic field,
${\bf B}=B_R \hat{\bf R} + B_\phi \hat{\rvecphi~}$,
 emanates from a perfectly conducting
star of radius $R_*$ and mass $M_*$ that rotates at the angular 
rate $\Omega_*$.
  It is assumed that $p={\rm const}~\rho^\gamma$
where $\gamma=$ const is the polytropic index.
  In the equatorial region the mass flux
of the wind per unit solid angle is conserved,
$\dot{M}_w/4\pi = R^2\rho v_R=$  const, and the 
radial magnetic flux per unit solid angle is also
conserved, $\Phi_R/4\pi = R^2 B_R=$ const.

   There are a number of conserved quantities of the flow,
one of these gives $B_\phi/B_R =(v_\phi -\Omega_* R)/v_R$,
another gives $L= R[v_\phi -B_R B_\phi/(4\pi v_R)]=$ const, the
angular momentum per unit mass, and a third gives the
Bernoulli constant ${\cal B}= (v_R^2+v_\phi^2)/2 +c_s^2/(\gamma-1)
-GM_*/R -\Omega_*R B_R B_\phi/(4\pi \rho v_R)=$ const, 
where $G$ is the gravitational constant and $c_s=(\gamma p/\rho)^{1/2}$
is the sound speed.   
   The ideal MHD equations can be combined to give an
equation for $d v_R/dR$ which can be integrated from
the surface of the star to very large distances 
for appropriately chosen parameters (BM76).
  Specifically, the parameters are chosen so as to allow
the smooth integration of $v_R$ through the slow and
fast magnetosonic singular points.

\begin{figure}
\includegraphics[scale=0.55]{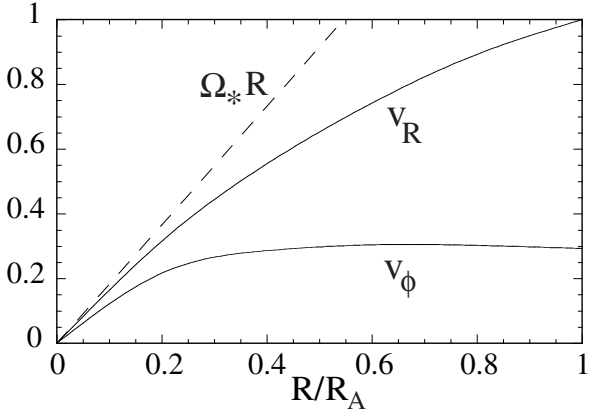}
\caption{Approximate radial ($v_R$) and azimuthal 
($v_\phi$) wind velocity components normalized
to $2v_M/3$ from M69.  The dashed line represents
the azimuthal corotation velocity. }
\end{figure}

    If the star is not rotating, the radial flow
velocity at very large distances is the Parker velocity 
$
v_P = [2 c_{s0}^2/  (\gamma-1) -2 G M_*/ R_*]^{1/2}~,
$
where $c_{s0}$ is the sound speed at the initial radius $R_*$
which is assumed larger than the flow speed at this distance.   
    For a rotating star, one can define the reference
velocity,
\begin{eqnarray}
v_M &\equiv& \left( { \Omega_*^2 (B_{R*}R_*^2)^2 
\over \dot{M}_w}\right)^{1/3}~,
\nonumber \\
&\approx& 1070 {{\rm km}\over {\rm s}}
\left({ 5~{\rm d} \over P_*}\right)^{2/3}
\left({B_{R*} \over 10^3 {\rm G}}\right)^{2/3}
\left({10^{-9} M_\odot/{\rm yr} \over \dot{M}_w}\right)^{1/3}
\end{eqnarray}
introduced by Michel (1969).  
Here, $P_*=2\pi/\Omega_*$ is the
rotation period of the star,  
$B_{R*} \equiv B_R(R_*)$, and in the second line
of equation (1) the star's radius is assumed
to be twice the radius
of the Sun, $R_*=1.4\times 10^{11}$ cm
(Armitage \& Clarke 1996).

\begin{figure}
\includegraphics[scale=0.55]{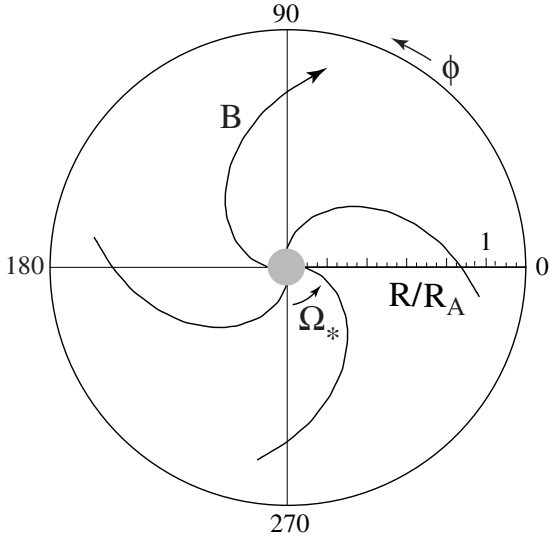}
\caption{Sample magnetic field lines
in the equatorial plane obtained using
equation (4). }
\end{figure}

     In this work we consider magneto-centrifugally
driven winds where $v_M \gg v_P$ and $v_M \gg c_{s0}$ 
as discussed by M69 and BM76.  
   In this limit
the radial wind velocity at large distances is  $v_M$.   
  The radial flow velocity increases from a
small value at the star  to a value
 $(2/3)v_M$ at the Alfv\'en radius 
\begin{eqnarray}
R_A &=& \left({3 \over 2}\right)^{1/2}{ v_M \over\Omega_*}~,
\nonumber\\
&\approx& 0.6 {\rm AU}
\left({  P_*\over 5~{\rm d}}\right)^{1/3}
\left({B_{R*} \over 10^3 {\rm G}}\right)^{2/3}
\left({10^{-9} M_\odot/{\rm yr} \over \dot{M}_w}\right)^{1/3}~
\end{eqnarray}
(M69).
   For $R \leq R_A$, approximations to the M69 solution are
\begin{equation}
v_R \approx {1.8 R/R_A\over 1+0.8R/R_A}~{2\over 3} v_M,~~~
v_\phi \approx {1.8 R/R_A\over (1+1.5R/R_A)^2}~{2\over 3} v_M~.
\end{equation}
These functions are shown in Figure 1.  
  Because $B_\phi/B_R =(v_\phi-\Omega_* R)/v_R$, it is clear from
equations (2) and (3) 
that this ratio  depends only on $R/R_A$.
  Figure 2 shows the shape of sample magnetic field lines in the
equatorial plane. 
   For $R=R_A$, $B_\phi/B_R \approx -1.55$.  
The ratio $\beta\equiv 4\pi\rho {\bf v}^2/{\bf B}^2$ also depends only
on $R/R_A$, and it increases with $R$ monotonically to 
$\approx 0.319$ at $R=R_A$.

\section{Drag Force on Planet}

  We consider a  giant planet of mass $M_p$ in a circular orbit
of radius $R_p \leq 1$ AU. 
   The geometry is sketched in Figure 3.
   The azimuthal drag force on the planet's motion through
the wind leads to a gradual change of the planet's angular
momentum $L_p=M_p(G M_* R_p)^{1/2}$:
\begin{eqnarray}
{d L_p \over dt} &=&{1\over 2}M_p\left({GM_* \over R_p}\right)^{1/2}
{dR_p \over dt}\equiv  {M_p(GM_*  R_p)^{1/2} \over 2T_w}
\nonumber\\
&=&{\rm sign}(v_\phi -v_{Kp}) R_p A_{\rm eff} P_{\rm ram}~.
\end{eqnarray}
Here, $T_w$ is the timescale for the wind 
drag to significantly change the
planet's orbit, $A_{\rm eff}$ is the effective cross section of
the planet discussed below, 
\begin{equation}
P_{\rm ram} = {1 \over 2} \rho (v_\phi -v_{Kp})^2 + {1\over 4\pi} {\bf B}^2~,
\end{equation}
is the ram pressure, and $v_{Kp}=(G M_*/R_p)^{1/2}$ is the Keplerian
velocity of the planet.

     We consider that the effective 
cross-section of the planet $A_{\rm eff}$
may be larger than its projected area $\pi r_p^2$
because the planet may have its own magnetic field or
the planet may have an evaporatively driven wind
(Vidal-Madjar, et al. 2003).
  We consider the case where the planet has a
dipole magnetic field with surface strength $B_{p0}$
(S\'anchez-Lavega 2004).
  Then, the effective area is
\begin{equation}
A_{\rm eff} = \pi r_p^2 \left({B_{p0}^2 \over 4\pi P_{\rm ram}(R_p)}
\right)^{1/3}~
\end{equation}
if this quantity is larger than $\pi r_p^2$;  otherwise
$A_{\rm eff}=\pi r_p^2$.

\begin{figure}
\includegraphics[scale=0.8]{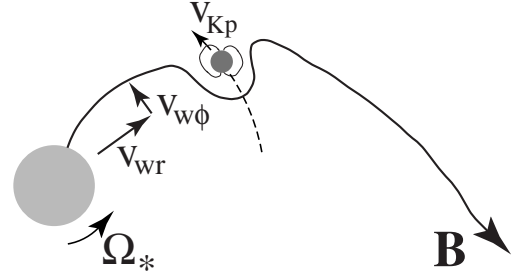}
\caption{Sketch of a magnetic field
line for conditions where the azimuthal
wind velocity $v_\phi$ is larger than
the planet's azimuthal velocity.}
\end{figure}

     The stellar wind causes 
{\it outward (inward)} migration of the
planet  when the azimuthal speed of the wind $v_\phi$  is 
{\it larger (smaller)}
than the planet's azimuthal speed $v_{Kp}$.
  We find that  equality $v_\phi=v_{Kp}$ holds 
for $R_p$ equal to a
critical value, $R_{cr}$, that depends  
on the parameters
of the wind, $\dot{M}_w,$ $B_{R*}$,  $v_M$, and the
star's rotation rate $\Omega_*=2\pi/P_*$.
    For
$R_{cr} \ll R_A$ we find
\begin{equation}
R_{cr} \sim \left({GM_*\over \Omega_*^2}\right)^{1/3}
\approx 0.87\times 10^{12}{\rm cm}\left({P_*\over5
{\rm d}}\right)^{2/3}~,
\end{equation}
which is the corotation radius of the star.
  A planet in a circular orbit of radius $R_{cr}$ has
an orbital period of $P_*$.
 This expression for $R_{cr}$ follows directly using
equation (3) which gives $v_\phi \approx 
\Omega_* R$ (for $R \ll R_A$)
and setting this equal to
the planet's Keplerian velocity $v_{Kp}$.

     Figure 4 shows the timescale $T_w$ from equation (4) for
the magnetic wind to cause a significant  change in the
planet's orbit  for different parameters.   
   For $R_p > R_{cr}$ ($R_p <R_{cr}$) the wind causes the planet to
move outward (inward).
    For this figure the planet's mass and radius have
been taken to be Jupiter's and its surface
magnetic field is taken to be $100$ G.  
    Typically, we find $A_{\rm eff} /\pi r_p^2 \sim 2 -10$.
 The torque on the planet acting to move it outward is
transmitted mainly by the magnetic field because we find
$R_{cr} \ll R_A$ where $\beta =4\pi \rho {\bf v}^2/{\bf B}^2 \ll 1.$
    The increase in the planet's angular momentum 
equals a corresponding decrease in the star's angular momentum. 
    Assuming $R_{cr} \ll R_A$ and using equation
(7) we find the estimate
\begin{eqnarray}
T_w &\sim& {2.06 M_p \over r_p^2}{(G M_*)^{11/9} \over \Omega_*^{13/9}
B_{p0}^{2/3} B_{R*}^{4/3} R_*^{8/3}}
\nonumber \\
&\approx& 17.5 {\rm Myr}\left({P_* \over 5 {\rm d}}\right)^{13/9}
\left({10^3 {\rm G} \over B_{R*}}\right)^{4/3}~,
\end{eqnarray}
where it is assumed that $A_{\rm eff} \geq \pi r_p^2$.
In the second line the planet's mass and radius is assumed equal
to Jupiter's and its surface magnetic field as $100$G.
The mass of the star is $M_\odot$ 
and its radius is $2R_\odot$.

\begin{figure}
\includegraphics[scale=0.45]{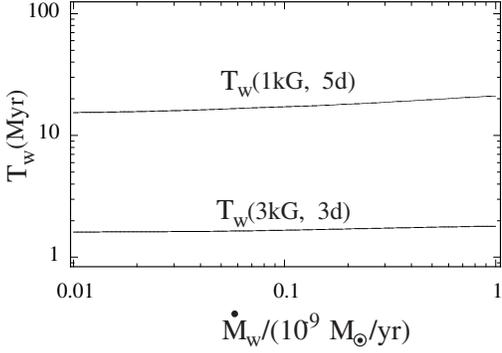}
\caption{Timescale for the wind
to significantly change the planet's
orbit $T_w$ as a function of the mass
loss rate of the wind, $\dot{M}_w$.  
  The bracketed quantities 
represent the star's surface magnetic
field in kG and its rotation period
in days.  The planet's mass and
radius are taken to be equal to 
Jupiter's while the planet's magnetic field
is taken to be $100$ G.
}
\end{figure}

     Alternatively, the torque on the planet can be
thought of in terms of a DC current circuit following
the model of Io's interaction with Jupiter's magnetosphere by
Goldreich and Lynden-Bell (1969).
  This is applicable if the Alfv\'en wave
transit time from the planet to the star and back is
less than the time for the planet to move a distance of
the order of its diameter relative to the wind
(Drell, Foley, \& Ruderman 1965).
   The torque is
 ${\cal T}=-R_p I_\theta B_R(R_p) (2 r_{\rm eff})/c$,
where $I_\theta$ is the poloidal current flowing in
the $\theta-$direction through planet's magnetopause.
   This current
then flows down the wind field line intersecting
the top of the planet at $\theta = r_{\rm eff}/R_p$;
it then flows through the star's atmosphere;  and 
finally it flows outward on the wind field line which
intersects the bottom of the planet at $\theta = -r_{\rm eff}/R_p$.
  The voltage drop across the planet in its reference
frame is $\Delta V = -(2r_{\rm eff})(v_\phi - v_{Kp})B_R(R_p)/c$,
which acts as a battery.  
   For representative values of $R_p=10^{12}$ cm, $B_R=20$ G, 
$r_{\rm eff}=7.15\times 10^9$ cm (Jupiter's radius), and
$v_\phi-v_{Kp}=10^6$ cm/s, one finds $\Delta V\sim 3\times 10^9$ V.
  The parts of the circuit along
the  field lines have negligible resistance.
  Thus the total current $I_\theta = \Delta V/{\cal R}$,
where ${\cal R}$ is the resistance of the parts of the circuit 
in the star's turbulent atmosphere and in the magnetopause
of the planet.   
      Setting the magnetic
field torque (equation 4) from the MHD model equal to the 
circuit model torque ${\cal T}$ gives the
total poloidal current flow, $I_\theta \sim 3.5 \times 10^{10}$ A 
for the mentioned conditions.  This implies that
the circuit resistance is ${\cal R} \sim 0.09$ Ohm.
The circuit picture is helpful for understanding the
interaction but a consistent treatment requires a 
treatment using the
MHD equations (Parker 1969).

    Observations and theory relevant to the 
 magnetic coupling of hot Jupiters to
their stars is discussed by
Cuntz, Saar, \& Musielak (2000), Shkolnik et al. (2005),
McIvor, Jardine, \& Holzwarth (2006), and Preusse et al.
(2006).

  { The stellar wind acts to spin-down the star 
giving a torque on the star 
$-\langle R_A^2 \sin^2(\theta)\rangle \Omega_*\dot{M}_w$
(Weber \& Davis 1967)
as discussed by Matt and Pudritz (2008a, b) for
the case of a thermally driven wind.
  Here, $\langle R_A^2 \sin^2(\theta)\rangle^{1/2}$ is
the mass-weighted, angle averaged  Alfv\'en radius 
for the angular  momentum outflow in the wind.   
    This average Alfv\'en radius may be significantly smaller than
the the values $R_A$ considered above.
 The angular momentum of the star is $k M_* R_*^2 \Omega_*$
with $k\approx 0.2$ for stars younger
than $10^7$ yr (Armitage \& Clarke 1996) so that the
spin-down time-scale of the star is 
\begin{equation}
T_{sd} \sim
{0.2M_* R_*^2 \over\dot{M}_w  \langle R_A^2 \sin^2\theta \rangle }~.
\end{equation}
For the reference parameters of equation (8) and $\dot{M}_w
\sim 10^{-9} - 10^{-10} M_\odot/{\rm yr}$, the spin-down time
$T_{sd}$ is less than the wind time $T_w$.  
Under such conditions the planet would at first 
have $R_p > R_{cr}$ and be pushed outward by the wind,
and subsequently it would have $R_p < R_{cr}$ and be pushed
inward.
}

\section{Influence of Tidal Interaction}

{  We consider that initially
the planet is at a relatively large distance with
$R_p > R_{cr}=(GM_*/\Omega_*^2)^{1/3}$
but with $R_p \ll R_A$. 
  We assume that the inner region of the disk ($<1$ AU) 
has been expelled so that there is no migration due
to the planet's interaction with the disk.

 For $R_p > R_{cr}$,  the rotating magnetic
wind tends  push the planet outward with a 
time-scale $T_w$ given by equation (8).   
   At the same time, the
magnetic wind causes the star to spin-down
with a time-scale $T_{sd}$ given by equation (9).
   If the spin-down time is longer than $T_w$ the planet
will be gradually pushed outward.  This case is
not considered further.
   In the opposite case where $T_{sd} < T_w$, 
the planet and star will evolve to have $R_p < R_{cr}$,
and the wind will tend to push the planet inward, again
with the time-scale of equation (8).
  However, the spin-down of the star acts to increase
$T_w$ and very likely $T_w/T_{sd}$ with the result that
the inward motion of the planet will cease.

  As the planet moves inward it 
raises an increasing tide on the star. 
 We assume that the planet itself is tidally locked
and in a circular orbit
(Marcy et al. 1997).  
    The time-scale for the establishment of 
corotation of the entire star
is estimated by Zahn (1994), Marcy et al. (1997), and
Donati et al. (2008) as
\begin{equation}
T_{tdl} \sim 1.2 \times 10^{12} 
{\rm yr}\left( {M_*\over 10^3 M_p}\right)^2 \left({R_p \over
10  R_*}\right)^6~,
\end{equation}
where   we have assumed a solar
mass star of radius $2R_\odot$.
The values of $T_{tdl}$ are probably longer than
the age of the systems except for very massive planets. 
 
  Marcy et al. (1997) and
Donati et al. (2008) suggest that only the convective
envelope of the star of mass 
$\delta M_* \ll M_*$ is brought into corotation
or ``locked'' to the planet.
  The time-scale for this to occur is reduced 
from  $T_{tdl}$ by a factor $\sim (\delta M_*/M_*)^2$. 
  The core of the star  then
rotates slower than its surface.
   For a Jupiter mass planet, $R_p =10^{12}$ cm, and $\delta M_*/M_* <
0.005$, we have $T_{tdl} < T_w$. 
  Of course, once the star's surface becomes locked to
the planet, the effect of the magnetic 
wind on the planet's orbit vanishes.
   With the planet/stellar surface locked, the loss of angular
momentum from the planet/steller surface system 
due to a magnetic wind  acts to reduce the
planet's period if it is initially relatively large.  
However, the reduction of the  angular 
momentum is limited for a fixed $\delta M_*$.
 This limit may such that a sufficiently
massive planet will fall into the star. 
For a less massive planet 
the system  behavior for further angular momentum loss is
unclear.

Note that the influence of the wind on the planet's orbit
decreases as $M_p$ increases and as the 
star's rotation period increases
(equation 8).  
On the other hand the tidal interaction increases
as $M_p$ increases, and it increases 
strongly as $R_p$ decreases (equation 10).
The remarkable object $\tau-$Bootis has a planet mass 
$M_p \approx 7.5M_J$   and period $P_p =3.3$ d  and is 
locked to the rotation of the star's surface (Donati et al. 2008;
Catala et al. 2007; Butler et al. 1997).  
  On the other
hand the famous object $51-$Pegasi (Mayor \& Queloz 1995) has a planet
of mass $ \approx 0.7M_J$ and a  period of $4.23$ d which orbits $51-$Peg
which has  a rotation period $P_*\approx 37$ d.  
  
   Consider what may happen to a
giant planet after its  fast inward
migration ends at a distance say $R_{pi} \lesssim 0.1$ AU 
owing to the planet entering a
magnetospheric gap or the  region of the disk
which has been blown away by a stellar wind.  
  The planet's distance $R_p$ may be unaffected by
the star's wind owing to the star's rapid spin down.
If the planet's mass is large enough relative to
the mass of the convective envelope, then it can cause 
tidal locking of the star's surface 
as observed in $\tau-$Boo.  For a smaller
mass planet there is no locking as is the case of  $51-$Peg.
  Another possibility is that $R_p$ changes due to
the influence of the magnetic wind.  
  The spin-down of the star is expected to give
$P_p < P_*$ in which case the wind causes a slow inward
motion of the planet or decrease of $R_p$.  
    The planet moves inward until there is 
tidal locking   with the star's surface 
(the time-scale for which is $\propto R_p^6$)   
   The tidal locking eliminates the wind affect
on the planet's orbit.  The migration due to 
the stellar wind allows smaller mass giant 
planets to produce tidal locking.

For tidal locking in a given period $T_{tdl}$, 
the distance where this occurs is
$R_p(tdl) \propto R_*(M_p/\delta M_*)^{1/3}$. 
For stars of decreasing mass $M_*< M_\odot$
(but $M_*>0.1M_\odot$), the star's radius decreases
strongly, and the mass of the convective envelope
$\delta M_*$ increases rapidly 
(the star is fully convective for $M_*< (0.3-0.4)M_\odot$)
(Chabrier \& Baraffe 2000).  Thus $R_p(tdl)$ decreases
rapidly as $M_p$ decreases from $M_\odot$.
}

\section{Erosion of the Disk by the Stellar Wind}

  {  Stellar winds
have long been considered as a mechanism for the dispersal
of accretion disks (Cameron 1973).  
   Theoretical and simulation studies of this 
process have been out assuming an
entirely hydrodynamic interaction (e.g., Cant\'o \& Raga 1991;
Richling \& Yorke 1997; Hollenbach, Yorke, \& Richstone 2000;
Soker 2005).
  Here we discuss the interaction of a high
velocity {\it magnetized} stellar wind
with the accretion disk.   The considered geometry is
shown in Figure 5.   The presence of the magnetic field leads
to Reynolds numbers which are sufficiently large that
the wind/disk boundary layer is strongly turbulent.  In
the hydrodynamic case the Reynolds numbers
are much smaller so that the boundary layer is probably
laminar (see Schlicting 1968, ch. 7). 
  In the following we derive an
estimate of the mass loss rate from the disk.}
    
   It is useful to consider the  physical conditions in the wind
at a distance $R=1$ AU for fiducial conditions 
of wind density $n_w=10^5$ cm$^{-3}$, wind speed
$v_w=500$ km/s, time averaged magnetic field $B_w=0.1$ G (predominantly
toroidal), and ion (proton) and electron temperatures $T_i=T_e = 10^5$ K.
  At this and larger distances note
that (i) the wind velocity is predominantly
radial, (ii) it is super fast magnetosonic, and (iii) it
is much larger than the Keplerian velocity of the disk matter.
  The ion and electron gyro-frequencies are $\omega_{ci}\approx
960$ s$^{-1}$ and $\omega_{ce}\approx 1.8\times 10^6$ s$^{-1}$;
the ion and electron gyro-radii are $r_{gi}\approx 3\times 10^3$ cm and
$r_{ge}\approx 70$ cm;  the ion and electron collision times
($\propto T^{3/2}/n$) are $\tau_i \approx 470$ s  and $\tau_e\approx 11$ s;
and the ion and electron mean-free paths are 
$\ell_i = \ell_e\approx 1.4 \times 10^9$ cm (Braginskii 1965).
   Thus we have $\omega_{ci}\tau_i \approx 4.6\times 10^5$
and $\omega_{ce}\tau_e\approx 2\times 10^7$.
   In the absence of a magnetic field, the kinematic viscosities 
of ions and electrons are $\nu_{0i}\approx v_{thi}^2\tau_i \approx
4\times 10^{15}$ cm$^2$ s$^{-1}$ and $\nu_{0e} \approx v_{the}^2 \tau_e
\approx 1.7\times 10^{17}$ cm$^2$ s$^{-1}$, 
where $v_{thi,e}$ are the ion or electron thermal speeds.
 Thus without a magnetic field the Reynolds number, 
$Re= R v_w/\nu_{0e} \approx 4400$, is such that a 
boundary layer flow is laminar. 
     With the magnetic field included, there are five different
viscosity coefficients.
   However, for the considered problem the
important viscosity coefficient is that for momentum transport
across the magnetic field; for example the
momentum flux-density component $\Pi_{R\theta}=-\rho \nu_\perp R^{-1} \partial
v_R/\partial \theta$.  
For the
ions it is $\nu_{\perp i} \approx \nu_{0i}/(\omega_{ci}\tau_i)^2
\approx 1.9\times 10^4 $ cm$^2$s$^{-1}$ and for electrons
$\nu_{\perp e} \approx \nu_{0e}/(\omega_{ce}\tau_e)^2
\approx 440 $ cm$^2$s$^{-1}$.  
    Using the viscosity $\nu_{\perp i}$, the effective
Reynolds number for the wind is 
$
Re_w = {R v_w / \nu_{\perp i}} \sim
5 \times 10^{17}$ so that the 
boundary layer flow is strongly turbulent.
This is in contrast with the non-magnetic case
where the boundary layer flow would be laminar.
     The large reduction of the viscosity
results from the particle step size
between collisions being a gyro-radius rather than
a mean-free path.  Thus the estimated Reynolds number
also holds for a turbulent magnetic field.

    Similarly, the important heat conductivity coefficient
is that for the heat flux across the magnetic field,
say $q_\theta =-\kappa_\perp R^{-1}\partial( k_B T)/\partial \theta$,
where $\kappa_\perp \approx \kappa_{\perp i} \approx 2 n v_{thi}^2 \tau_i
/(\omega_{ci}\tau_i)^2 \approx 3.8 \times 10^9$ $({\rm cm~s})^{-1}$
and where $k_B$ is Boltzman's constant.
   For this heat conductivity, the heat flow from
the wind into the disk is negligible compared with
the energy flux per unit area from the disk $3GM_*\dot{M}_a/(8\pi R^3)$
for accretion rates 
$\dot{M}_a = 10^{-10} - 10^{-8} M_\odot$yr$^{-1}$.

\begin{figure}
\includegraphics[scale=1.]{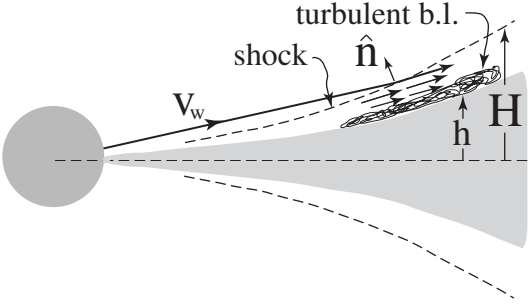}
\caption{Sketch of  disk-wind
turbulent boundary
layer with the outer oblique shock.
$\hat{\bf n}$ is the normal to the
shock wave, ${\bf v}_w$ is the wind velocity,
$h$ is the disk half-thickness, and $H$ is
the vertical height of the shock.
}
\end{figure}

   At the place where the wind encounters 
the much denser disk,
there is a weak oblique shock at a height
$H(R) \ll R$ above the equatorial plane
and above the disk which has a half-thickness
$h(R)<H(R)$ as sketched in Figure 5. 
   The angle between the incident flow and the shock is
$\beta \approx R d(H/R)/dR \ll 1.$  
  In general we expect to have $\beta >0$.
  In passing through the shock
the flow is deflected through an angle $\delta =2\beta/(1+\gamma)
=3\beta/4$ (for $\gamma=5/3$) away from the equatorial plane,
and the  flow speed is reduced by a small fractional amount.
  The region between $h$ and $H$ is referred to as the
boundary layer.  The influx of wind matter into this
layer is $-d{\bf S} \cdot(\rho_w 
{\bf v}_w)\approx dS\rho_w v_w R[d(H/R)/dR]$,
where $dS \approx RdR d\phi$ is the area element of the  shock on the
top side of the disk.  
  The Keplerian velocity of the disk is
small compared with the wind velocity for
$R\geq 1$ AU, and it
is neglected.   
  The density in the boundary layer varies
from $\rho(R,h) \gg \rho_w$ at the surface of
the disk to $\rho_w$ at $z=H$.
  The time-averaged radial flow velocity varies from $| v_R(R,h) | \ll
v_w$ to $v_R  =v_w$ at $z=H$.  Here, we neglect $v_R(R,h)$.
    
For stationary conditions, the conservation of mass and
radial momentum
in the annular region $abcd$ of the boundary layer on the
top side of the disk - sketched
in Figure 6 - gives
\begin{eqnarray}
{\partial \over \partial R} (R F^m_R) &=& 
R\left[R {d(H/R)\over dR}\right]\rho_w v_w
+{1\over 2\pi}{d\dot{M}_d \over dR}\\
{\partial \over \partial R} (R F^p_R) &=& 
R\left[R {d(H/R)\over dR}\right]\rho_w v_w^2~,
\end{eqnarray} 
where the term on the left-hand side is from the vertical sides
of the region and the right-hand side is due to the sides $ab$
and $cd$.
  Here, $F_R^m$ is the mass flux and $F_R^p$ is the radial momentum
flux both per unit circumference of the top side
of the disk.  That is,
\begin{equation} 
F_R^m=\int_h^H dz~ \langle \rho  v_R \rangle~,
\quad F_R^p = \int_h^H dz ~ 
\langle\rho  v_R^2 \rangle~,
\end{equation}
where the averages are over the turbulent fluctuations.
The mass loss rate of disk  per
unit radius due to entrainment is $d\dot{M}_d/dR$.   The
disk matter influx to the boundary
layer brings in negligible radial momentum.  

    We now integrate equations (11) and (12) 
from an inner radius $R_{\rm in}$, where
$F_R^p$ and $F_R^m$ are negligible, to distance $R$.
We define an average radial velocity
in the boundary layer as $u(R) \equiv 
\int dz \langle \rho v_R^2\rangle 
 /\int dz \langle \rho v_R \rangle $. 
Thus we obtain an equation for the mass loss rate
from the top and bottom 
sides of the  disk between $R=R_{\rm in}$ and $R$,
\begin{equation}
\dot{M}_d =4\pi\left({v_w \over u}-1\right)\int_{R_{\rm in}}^R
R dR \left[ R{d(H/R) \over dR}\right] \rho_w v_w~,
\end{equation}
which is von Karman momentum integral for this problem
(Schlicting 1968, ch. 8).
Using the fact that
$\rho_w v_w =\dot{M}_w/(4\pi R^2)$
we can evaluate this
integral at the outer radius of the disk $R_{\rm
out}$.  This gives the mass loss rate of the  disk,
\begin{equation}
\dot{M}_d =\left({v_w \over u}-1\right)_{\rm out}
\left({H \over R}\right)_{\rm out} \dot{M}_w~.
\end{equation}
The main unknown in this equation is $u$.

The average velocity
$u$ depends on the vertical 
profiles of density and radial
velocity which are not known.
   We expect the profiles, for example
$\langle v_R(z)\rangle$ to be
substantially different from those of
laboratory turbulent boundary layers
over solid surfaces (e.g., Schlicting  1968, ch. 23;
Roy \& Blottner 2006).
The main reason for the difference is
that the  density at the
surface of the disk $\rho(R,h)$
is many orders of magnitude larger 
than the wind density $\rho_w$.
  For a laboratory
boundary layer, a mixing-length model of the
momentum transport gives $(z^\prime)^2
(d\langle v_R\rangle/dz^\prime)^2=$ const
(with $z^\prime \equiv z-h$) , and this gives
the well-known logarithmic velocity profile
(see e.g. Schlicting 1968).
  For this profile most of the change
of velocity is quite close to the wall ($z^\prime =0$).
  In contrast, for the disk boundary
layer  a mixing
length model  gives $\rho(z^\prime)(z^\prime)^2
(d\langle v_R\rangle/dz^\prime)^2=$ const.
Because of the density dependence, the change in
the velocity  occurs  relatively far 
from the wall.
   An important consequence of this is that
the average velocity $u$ is much smaller
than $v_w$.  
   If we make the weak
assumption that $u$ is less than $v_w/2$, 
then we have a lower limit on the disk
mass loss rate $\dot{M}_d \gtrsim
(H/R)_{\rm out} \dot{M}_w$.  
  For an initial disk mass of $M_d=0.02 M_\odot$
(e.g., Kuchner 2004),
$(H/R)_{\rm out}=0.2$, and $\dot{M}_w =10^{-9}
M_\odot$yr$^{-1}$, the disk loss time
is  $T_d\lesssim M_d/\dot{M}_d =10^8$ yr.

    However, the abovementioned discussion
suggests that $u$ is much less than $v_w$.
Thus the above upper limit on $T_d$ is a
large over estimate of the erosion time.
 Because $u$
is necessarily larger
than the local escape velocity 
$v_{\rm esc}=(2GM_*/R)^{1/2}$
in order for the  matter flow in the boundary
layer to become unbound from the star.
   Thus we find an upper bound on
the mass loss rate from the disk due to the
wind
\begin{eqnarray}
\dot{M}_d &\lesssim &\left( { v_w H \over 
v_{\rm esc} R }\right)_{\rm out} \dot{M}_w~,
\nonumber\\
&\lesssim& 17\!\left({v_w \over 500{\rm km/s}}\right)\!
\left({R_{\rm out} \over 50{\rm AU}}\right)^{1/2}\!
\left({(H/R)_{\rm out} \over 0.2}\right) \dot{M}_w,\quad
\end{eqnarray}
assuming $v_{esc}\ll v_w$.
  For an initial disk mass of $M_d=0.02 M_\odot$,
$(H/R)_{\rm out}=0.2$, $R_{\rm out}=50$ AU, 
and $\dot{M}_w=10^{-9} M_\odot$yr$^{-1}$, 
the disk loss time is  
$T_d\gtrsim M_d/\dot{M}_d =1.2\times 10^6$ yr.
   Equation (16) has a different dependence
on $H/R$ and gives different values compared
with the hydrodynamic estimates (Hollenbach
et al. 2000).

\begin{figure}
\includegraphics[scale=1.1]{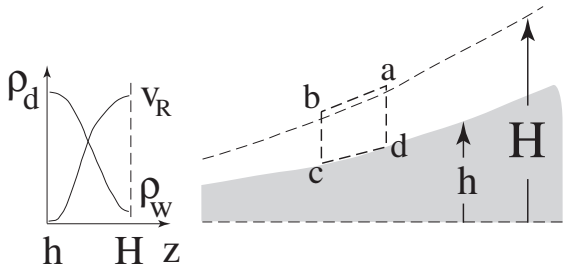}
\caption{ The right-hand
part of the figure shows the control
volume $abcd$ discussed in the text. 
  The left-hand side of the figure shows
the vertical profiles of the
density $\rho$ and radial velocity $v_R$
in the boundary layer. Here, $\rho_d$
is the density at the surface of the disk,
and at $z=H$, $v_R=v_w$.
}
\end{figure}

\section{Conclusions}

   We conclude that 
magneto-centrifugally driven winds of strongly magnetized, 
rapidly rotating T Tauri stars 
may affect the orbits of close-in giant planets that
are inside the Alfv\'en radius of the wind. 
   The magnetic wind may of course  disperse the inner part
of the disk and thereby halt the inward migration of
a giant planet.  
   However, this work focuses on the affect of 
the azimuthal ram pressure of the magnetized wind
on the planet.   This gives a torque which tends
to increase (decrease) the planet's angular momentum 
thereby causing it to move outward (inward)
if the planet is at a distance $R$ larger (smaller)
than the corotation radius of the star,
$R_{cr} =(GM_*/\Omega_*^2)^{1/3}\approx
0.07 {\rm AU}(P_*/5{\rm d})^{2/3}$ for a solar mass
star, where $\Omega_*=2\pi/P_*$ is the star's angular rotation
rate. 
    Such winds are likely to occur in T Tauri
stars where the thermal
speed of the gas close to the star is small
compared with the escape velocity,
where the star's magnetic field
is strong, and where the star rotates rapidly. 
   The magnetic winds 
are expected to be important
for giant planet migration for  cases where the star's
magnetic field  is a multipolar field rather dipolar.
  For an approximately
aligned dipolar magnetic field  the
wind is directed away from the
equatorial plane and  a magnetospheric
cavity forms. (Lin et al. 1996;  
Romanova \& Lovelace 2006).
    We find that the time-scale for  the magnetic wind to change
the planet's orbit $T_w$ ranges from $2 -20$ Myr for stellar
rotation periods of $3-5$ d and surface magnetic fields of $1-3$
kG. This time-scale is of the order of or longer than the
estimated spin down time of the star $T_{sd}$ due to the magnetic
wind.

   The migration of a close in giant planet due to the
star's wind  may be
strongly affected by the tidal interaction.   
The time-scale for the planet to establish tidal locking $T_{tdl}$
of the entire star is estimated to be longer than the
age of most systems.  However, locking of the convective
envelope of the star may occur on a much shorter time-scale.
This time scale varies as the sixth power of the distance
to the planet.  Once locking of the stars surface 
occurs, the influence of the wind on the planet's
orbit vanishes.

    Compared with the magnetospheric cavity
model proposed for halting the inward migration
of planets (Lin et al. 1996; Romanova \& Lovelace
2006), the influence of the magnetic wind may allow for
a broader distribution of distances of exosolar
planets (Pont 2007).

    The influence of the magnetic wind
in eroding and eventually destroying the accretion disk is
analyzed.    A
momentum integral for the boundary layer is derived and used
to estimate the disk erosion time-scale $T_d$. 
   For the considered conditions we find
 $T_d \sim 1-10^2$ Myr with
the lower value  favored.

We thank Ralph Pudritz for valuable criticism
of an earlier version of this work.  
This work was supported in
part by NASA grants NAG5-13220 and
NAG5-13060 and by NSF grant AST-0507760.

\end{document}